\documentclass[%
 reprint,
 amsmath,amssymb,
 aps,
 prl,
]{revtex4-1}

\usepackage[utf8]{inputenc}
\usepackage{graphicx}
\usepackage{dcolumn}
\usepackage{bm}
\usepackage[colorlinks=true,linktocpage=true,linkcolor=blue,citecolor=blue,allcolors=blue]{hyperref}

\usepackage{color}
\usepackage{xspace}
\usepackage[capitalize]{cleveref}

\newcommand{\dd}{{\rm d}}

\newcommand{\UNIT}[1]{\ensuremath{\,{\rm #1}}\xspace}

\newcommand{\MeV}{\UNIT{MeV}}

\newcommand{\fm}{\UNIT{fm}}
\newcommand{\fmc}{\ensuremath{\,{\rm fm}/c}\xspace}

\newcommand{\proz}{\UNIT{\%}}

\newcommand{\mb}{\UNIT{mb}}

\newcommand{\REM}[1]{}

\definecolor{magenta}{cmyk}{0,1,0,0}

\newcommand{\pluseq}{\mathrel{+}=}

\newcommand*{\defeq}{\mathrel{\vcenter{\baselineskip0.5ex \lineskiplimit0pt
                     \hbox{\scriptsize.}\hbox{\scriptsize.}}}%
                     =}

\newcommand{\all}[2]{\alpha_{_{{#1}\rightleftharpoons{#2}}}}
\newcommand{\al}[2]{\tilde\alpha_{_{{#1}\rightleftharpoons{#2}}}}

\newcommand{\cR}[2]{c_{#1}^{#2}}

\newcommand{\ki}{K}
\newcommand{\ka}{\overline{K}}

\newcommand{\nA}{\rm A}
\newcommand{\nD}{\rm d}
\newcommand{\nT}{\rm t}

\begin{document}


\title{Solving the puzzle of high temperature light (anti)-nuclei production in ultra-relativistic heavy ion collisions}

\author{Tim Neidig}
\thanks{Corresponding author}
\email{neidig@itp.uni-frankfurt.de}
\affiliation{Institut f\"ur Theoretische Physik, Johann Wolfgang Goethe-Universit\"at, Max-von-Laue-Strasse 1, 60438 Frankfurt am Main, Germany}%

\author{Kai Gallmeister}
\altaffiliation[Present address: ]{Institut f\"ur Theoretische Physik, Justus-Liebig-Universität, Heinrich-Buff-Ring 16, 35392 Gießen, Germany}
\affiliation{Institut f\"ur Theoretische Physik, Johann Wolfgang Goethe-Universit\"at, Max-von-Laue-Strasse 1, 60438 Frankfurt am Main, Germany}%

\author{Carsten Greiner}
\affiliation{Institut f\"ur Theoretische Physik, Johann Wolfgang Goethe-Universit\"at, Max-von-Laue-Strasse 1, 60438 Frankfurt am Main, Germany}%

\author{Marcus Bleicher}
\affiliation{Institut f\"ur Theoretische Physik, Johann Wolfgang Goethe-Universit\"at, Max-von-Laue-Strasse 1, 60438 Frankfurt am Main, Germany}%
\affiliation{Helmholtz Research Academy Hesse for FAIR (HFHF), GSI Helmholtz Center, Campus Frankfurt, Max-von-Laue-Stra{\ss}e 12, 60438 Frankfurt am Main, Germany}

\author{Volodymyr Vovchenko}
\affiliation{Nuclear Science Division, Lawrence Berkeley National Laboratory, 1 Cyclotron Road,  Berkeley, CA 94720, USA}

\date{\today}

\begin{abstract}
The creation of loosely bound objects in heavy ion collisions, e.g.~light clusters, near the phase transition temperature ($T_{\rm ch} \approx 155\MeV$) has been a puzzling observation that seems to be at odds with Big Bang nucleosynthesis suggesting that deuterons and other clusters are formed only below a temperature $T\approx 0.1-1\MeV$. We solve this puzzle by showing that the light cluster abundancies in heavy ion reactions stay approximately constant from chemical freeze-out to kinetic freeze-out. To this aim we develop an extensive network of coupled reaction rate equations including stable hadrons and hadronic resonances to describe the temporal evolution
of the abundancies of light (anti-)(hyper-)nuclei in the late hadronic 
environment of an ultrarelativistic heavy ion collision.
It is demonstrated that the chemical equilibration of the light 
nuclei occurs on a very short timescale as a consequence of the strong 
production and dissociation processes. 
However, because of the partial chemical equilibrium of the stable hadrons, including the nucleon feeding from $\Delta$ resonances, the abundancies of the light nuclei stay nearly constant during the evolution and cooling of the hadronic phase. This solves the longstanding contradiction between the thermal fits and the late stage coalescence (and the Big Bang nucleosynthesis) and explains why the observed light cluster yields are compatible with both a high chemical production temperature and a late state emission as modelled by coalescence. We also note in passing that the abundancies of the light clusters in the present approach are in excellent agreement with those measured by ALICE at LHC.
\end{abstract}

\maketitle


The production of light nuclei in an expanding hadronic medium has been the driving force of the Big Bang nucleosynthesis that has led to the creation of the light atomic nuclei in the universe. Recently, similar conditions were recreated in ultra-relativistic heavy ion collisions to allow for a deeper understanding of light cluster production under controlled collisions. 
Especially, the yields of light nuclei like deuterons, tritons, helium-3 and helium-4, their anti-particles, and also hyper-tritons, have been measured by the ALICE collaboration at LHC \cite{ALICE:2015oer, ALICE:2015wav, ALICE:2019fee}. It was found that their abundances are in a remarkable agreement with the predictions of the statistical hadronization model, characterized by the chemical freeze-out temperature of $T_{\rm ch} = 155 \MeV $ and  nearly vanishing net baryon density~\cite{Andronic:2010qu,Andronic:2017pug}.
This was a surprising result, because one would usually expect (in line with Big Bang nucleosynthesis) that the production of such loosely bound states with binding energies of ${\cal O}(2\MeV)$ would only happen at much lower temperatures on the order of the binding energy. The idea of late, i.e. low temperature cluster formation is usually realized via the 
coalescence of nucleons in the final state
\cite{Mrowczynski:2016xqm,Sombun:2018yqh,Zhao:2018lyf,Bellini:2020cbj, Glassel:2021rod} or through kinetic transport approach describing the continuous production and dissociation of such light nuclei \cite{Oliinychenko:2018ugs, Oliinychenko:2020znl, Sun:2021dlz, Staudenmaier:2021lrg}. 
The two latter descriptions have certain shortcomings, for instance, the energy conservation is not obeyed in the coalescence approach whereas the former treatment of nuclei as point-like particles in the kinetic approach may be questionable, but recent works are going beyond this treatment e.g. \cite{Sun:2021dlz}. At first glance, both approaches seem incompatible with each other, however, as we will show below this is not the case (see also \cite{Mrowczynski:2016xqm} for an alternative discussion of this topic). 

Our explanation is based on the assumption of detailed balance -- the law of mass action. 
Recently, this principle was used to show how the final yields of light nuclei are determined 
by the initial (chemical) freeze-out multiplicities of the stable hadrons, akin with the nuclear equivalent of the Saha ionization equation of cosmology \cite{Vovchenko:2019aoz}. In \cite{Gallmeister:2020fiv}, this idea of detailed balance resp.~a Saha equation has also been studied in a picture, where the usual hadron gas prescription has been extended by so called Hagedorn resonances \cite{Gallmeister:2020fiv}. 

In a heavy-ion setup one uses the framework of partial chemical equilibrium (PCE) \cite{Bebie:1991ij} to describe the hadronic environment after the chemical freeze-out. 
The total abundances of stable hadrons are fixed by introducing non-equilibrium chemical potentials~\cite{Xu:2018jff,Vovchenko:2019aoz}, while the yields of resonances change~\cite{Motornenko:2019jha,Tomasik:2021jfd}, being determined by the relative equilibrium of decays and regenerations. The light nuclei are assumed to be in relative chemical equilibrium in this thermal bath.

In this letter we relax the assumption of instantaneous equilibration of nuclear reactions, as well as resonance decays and regenerations, by means of a 
set of coupled (reaction) rate equations describing the chemical composition of stable hadrons (pions, kaons, (anti-)nucleons, $\Lambda $s),  hadronic resonances and light (strange) (anti-)nuclei. 
Such a description of the
chemical evolution of the degrees of freedom and especially also of rare probes is well known since a long time in relativistic heavy ion collisions (e.g.~(multi-) strange baryons \cite{Koch:1986ud, Barz:1988md, Noronha-Hostler:2007fzh, Noronha-Hostler:2009wof}) and in cosmology (e.g.~nucleosynthesis, see \cite{Kolb:1990vq} and references therein).


The particles of special interest in the rate equations are the light nuclei ${\rm d}(={\rm H}^2)$, ${\rm t}(={\rm H}^3)$, ${\rm He}^3$, ${\rm He}^4$ and the strange light nucleus ${\rm H}_\Lambda^3$ (and their antiparticles). 
In the mesonic sector, $\pi$, $K$ and $\overline K$ are considered as stable particles, while additionally the resonances $\rho$, $\omega$ and  $K^*$ and $\overline{K^*}$ are included.
The baryonic sector is covered by nucleons $N(=n,p)$, the $\Delta(1232)$ resonance and the strange $\Lambda$ baryon as the dominating species. Also here all antiparticles are taken into account.
For simplicity, we omit higher lying $\Delta$ and $N^*$ resonances~\footnote{Essentially the same evolution of particle multiplicities is obtained when using the full particle list, as implemented within the \texttt{Thermal-FIST} package~\cite{Vovchenko:2019pjl} and reported in Ref.~\cite{VovchenkoTalkCERN2021}.}. In order to compensate this omission, an effective degeneracy for the $\Delta(1232)$ resonances is introduced, which has been chosen to be twice the fundamental degeneracy, $g_\Delta^{\rm eff}=2g_\Delta$, in order to fit the initial multiplicities  in accordance to the statistical hadronisation model~\cite{Andronic:2017pug}. 

In the rate equation approach two different types of reactions are considered: first the catalized break up (and fusion) reactions for the nuclei 
\begin{align}
A + X \rightleftharpoons a \cdot N + X
\label{eq:45}
\end{align}
and secondly the decay and formation of unstable resonances
\begin{align}
R \rightleftharpoons X + Y\ .
\label{eq:46}
\end{align}
As all the reactions are of the type $(2\leftrightarrow n)$, the collision rates can be expressed in terms of binary cross sections.

In the case of resonance decays, the change in the multiplicity of resonance $R$ with time $t$~\footnote{The time can be interpreted as the comoving Bjorken time} may be expressed as e.g. for two decay particles
\begin{align}
\label{eq:reso_rate}
  \frac{\dd N_R}{\dd t}
  = -\all{R}{X+Y}(N_R-\cR{R}{XY}N_XN_Y)\,
\end{align}
where $\all{R}{X+Y}$ can be interpreted as the decay rate of $R\to X+Y$, but also as a scaled cross section for $X+Y\to R$. The factor $\cR{R}{XY}\defeq N_R^{\rm eq}/(N_X^{\rm eq}N_Y^{\rm eq})$, given by the ratio of the multiplicities in equilibrium, $N_i^{\rm eq}$, is dictated by the detailed balance. Any $N_i^{\rm eq}$ is calculated at full thermal and chemical equilibrium.
The contributions of resonance $R$ decays and regenerations to the change rate of the multiplicities of the decay products, $\dd N_X/\dd t$ and $\dd N_Y/\dd t$ are given by the rhs. of Eq.~\eqref{eq:reso_rate} with the opposite sign.

Considering the nuclear reactions, catalized by some particle (e.g.~pionic~\cite{Oliinychenko:2018ugs}), similar expressions can be written down, based on the knowledge of the basic $2\to n$ process.

The decay rates used are thermal averaged values over experimentally measured cross sections (e.g.~$\sigma^{\rm{inelastic}}_{\pi+\rm{d}}$) or resonance decay widths taken from Particle Data Tables~\cite{ParticleDataGroup:2020ssz},
\begin{align}
\begin{split}
\alpha_{_{A+X \rightarrow a N + X}} 
&= \frac{\bigl\langle \sigma_{_{A+X \rightarrow a N + X}} v_{\rm rel}\bigr\rangle}{V} N_{X} \\
&= \tilde{\alpha}_{_{A+X \rightarrow a N + X}} N_{X}\ .
\end{split}
\end{align}
Here we introduced the numerical factor $\tilde{\alpha}$ 
to stress the explicit dependence of $\alpha$ of the catalyzing mesons, whose multiplicities are also time dependent, as they are part of the network.

For the light nuclei we approximate the cross sections by summing over their constituents, e.g.~$\sigma_{\nT+\pi \rightarrow 3 N + \pi} \approx 2 \sigma_{_{\nD+\pi \rightarrow 2 N + \pi}} - \sigma_{{p+\pi \rightarrow p + \pi}}$.
Since experimental data for the cross section for $\pi/K + {\rm H}_{\Lambda}^{3}$ are not available, the values of $90\mb$ resp.~$30\mb$ are used, based on the reaction $\pi/K + {\rm H}^{3}$.

In order to calculate the multiplicities of particles created in heavy ion collisions, 
we incorporate the expansion and the cooling of the fireball in the hadronic phase, through a time dependence of the volume, $V(t)$, and the temperature, $T(t)$.
For a given (expanding) volume $V$ the temperature $T$ is obtained by in the framework of the PCE, so that the abundance of each stable hadron species (including the resonance contribution) are conserved, and by conserving the total entropy of the system as well as the net baryon number and net strangeness.
These relations provide a set of non-linear equations, which can be solved for any $T \le T_{\rm ch}$ to obtain $V(T)$ and $\mu_{i}(T)$ ($i=S,B,N,\pi$ etc.). The initial conditions at $T_{\rm ch}$ namely $V(T_{\rm ch})$ and the baryon and strangeness chemical potential are obtained from the experimental data \cite{ALICE:2015oer, ALICE:2015wav, ALICE:2019fee, Abelev:2013vea, Abelev:2013xaa, Andronic:2017pug}.
Finally, the time dependence of the volume is given by the parametrization that incorporates longitudinal and transverse expansion \cite{Pan:2014caa},
\begin{align}
    V(t)= V_{\rm ch} \frac{t}{t_{\rm ch}} \frac{t_{\perp}^{2}+t^{2}}{t_{\perp}^{2}+t_{\rm ch}^{2}}
\label{eq:V}
\end{align}
with $t_{\perp}=6.5 \fmc$ and $t_{\rm ch}=9\fmc$.
In the present work, all calculations assume an initial temperature $T_{\rm ch}=155 \MeV$ for the chemical freeze-out.

In total, the network contains 23 rate equations, given by the number of stable and unstable particles and antiparticles. In the following, we will only list the most important ones.

As a first example, the rate equation for the deuteron is given as
\begin{align}
    \frac{\dd N_{\nD}}{\dd t}
    =& -\sum_{x=\pi,\ki,\ka} \al{\nD+x}{2N+x} N_{x} (N_{\nD} - \cR{\nD}{N^2} \, N_{N}^{2})
    \label{eq:6}
\end{align}
with the generalized definition
$\cR{AB\dots}{XY\dots}\defeq \frac{N_A^{\rm eq}N_B^{\rm eq}\cdots}{N_X^{\rm eq}N_Y^{\rm eq}\cdots}$ and
using shortcuts like $\cR{(\cdots)}{XX}=\cR{(\cdots)}{X^2}$, 
while the rate equation for the nucleons is given by

\begin{align}
  \begin{split}
    &\frac{\dd N_N}{\dd t}
    = \al{\Delta}{N+\pi} ( N_{\Delta} - \cR{\Delta}{N\pi} \, N_{N} N_{\pi})
    +\sum_{X=\nD,\nT,{\rm He}^{3},{\rm He}^{4}} A_X  \\
    &\ \times \sum_{x=\pi,\ki,\ka} 
    \al{X+x}{A_X+x} N_{x} ( N_X - \cR{X}{N^{A_X}} \, N_{N}^{A_X})\\
    &\ +\sum_{x=\pi,\ki,\ka}2\al{{\rm H}_{\Lambda}^{3}+x}{NN\Lambda+x} N_{x} ( N_{{\rm H}_{\Lambda}^{3}} - \cR{{\rm H}_\Lambda^{3}}{N^2\Lambda} \, N_{N}^{2} N_{\Lambda}).
\end{split}
\label{eq:7}
\end{align}

Here $A_X$ is the nucleon content of a nucleus $X$.
The rate equations for pions, being a catalyzing particle in many equations, is itself quite simple, since its
affected only by the resonances $\Delta$, $\rho$ and $\omega$,
\begin{align}
  \begin{split}
  \frac{\dd N_\pi}{\dd t}
  =&\ \ \al{\Delta}{N\pi}(N_{\Delta}-\cR{\Delta}{N\pi}N_NN_\pi)\\
  & + \al{\overline{\Delta}}{\overline{N}\pi}(N_{\overline{\Delta}}-\cR{\overline{\Delta}}{{\overline{N}\pi}}N_{\overline{N}}N_\pi)\\
  & + 2\al{\rho}{2\pi}(N_\rho-\cR{\rho}{\pi^2}N_\pi^2) \\
  & + 3\al{\omega}{3\pi}(N_\omega-\cR{\omega}{\pi^3}N_\pi^3)\ .
  \end{split}
  \label{eq:8}
\end{align}

The chemical processes described via \cref{eq:7,eq:8}, and also the others conserve the total number of stable hadrons, either direct or carried in a resonance or light nuclei. When the rates are indeed large compared to the expansion rate, then the abundancies of the stable hadrons fulfill PCE.

\begin{figure}[htb]
  \begin{center}
    \hspace*{\fill}%
    \includegraphics[width=0.95\columnwidth,clip=true]{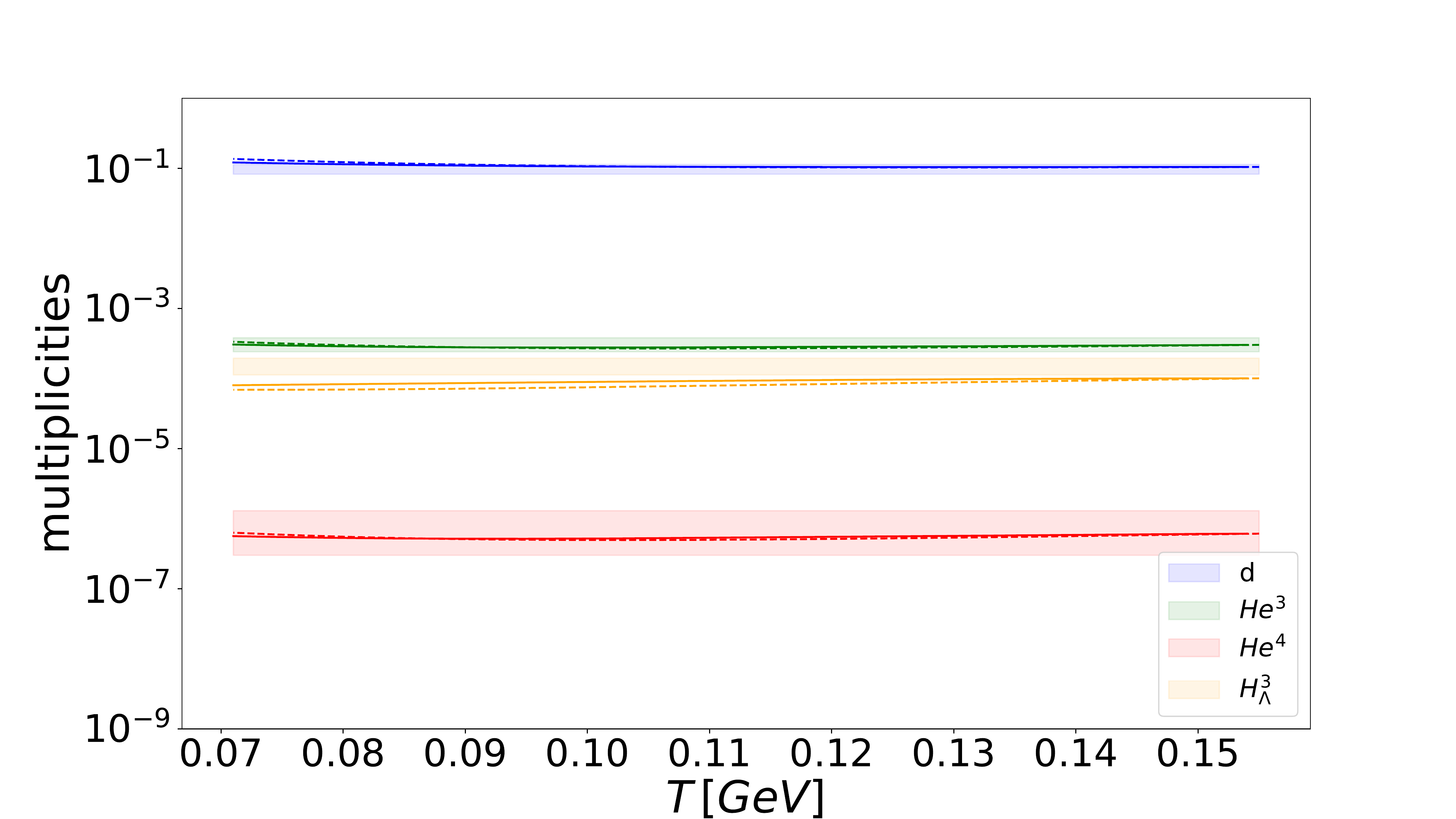}
    \hspace*{\fill}%

    \caption{
      The multiplicity of the light nuclei as function of the (decreasing) temperature $T$. Solid lines represent the results of the rate equations, while dashed curves show the result of the Saha equation. The colored bands represent the experimental data (ALICE) \cite{ALICE:2015oer, ALICE:2015wav, ALICE:2019fee} with their uncertainty.
    }
    \label{fig:yields}
  \end{center}
\end{figure}

In \cref{fig:yields} we show the results for the yields of the light nuclei using the rate equations. 
These results are compared to the ones obtained in the Saha equation limit~\cite{Vovchenko:2019aoz}, as well as to the experimental data of the ALICE Collaboration~\cite{ALICE:2015oer, ALICE:2015wav, ALICE:2019fee}. The calculation starts at the chemical freeze-out, $T_{\rm ch}=155 \MeV$, and covers temperatures down to $T = 70\MeV$.
The yields of light nuclei exhibit only a very small temperature dependence \cite{Xu:2018jff},
with a minor exception of ${\rm H}_\Lambda^3$, which
is attributed to the missing hyperon resonances~\cite{Vovchenko:2019aoz}.
Generally, a very good agreement with the experimental data is obtained for the whole temperature range. This indicates that one can not distinguish the emission of light clusters from the chemical freeze-out hypersurface from an emission during the later stages of the reaction by looking at the multiplicities alone. Let us remark in passing that the rate equations results stay within 5-10\proz of the Saha equation in the whole temperature range, supporting our previous findings.

\begin{figure}[htb]
  \begin{center}
    \hspace*{\fill}%
    \includegraphics[width=0.95\columnwidth,clip=true]{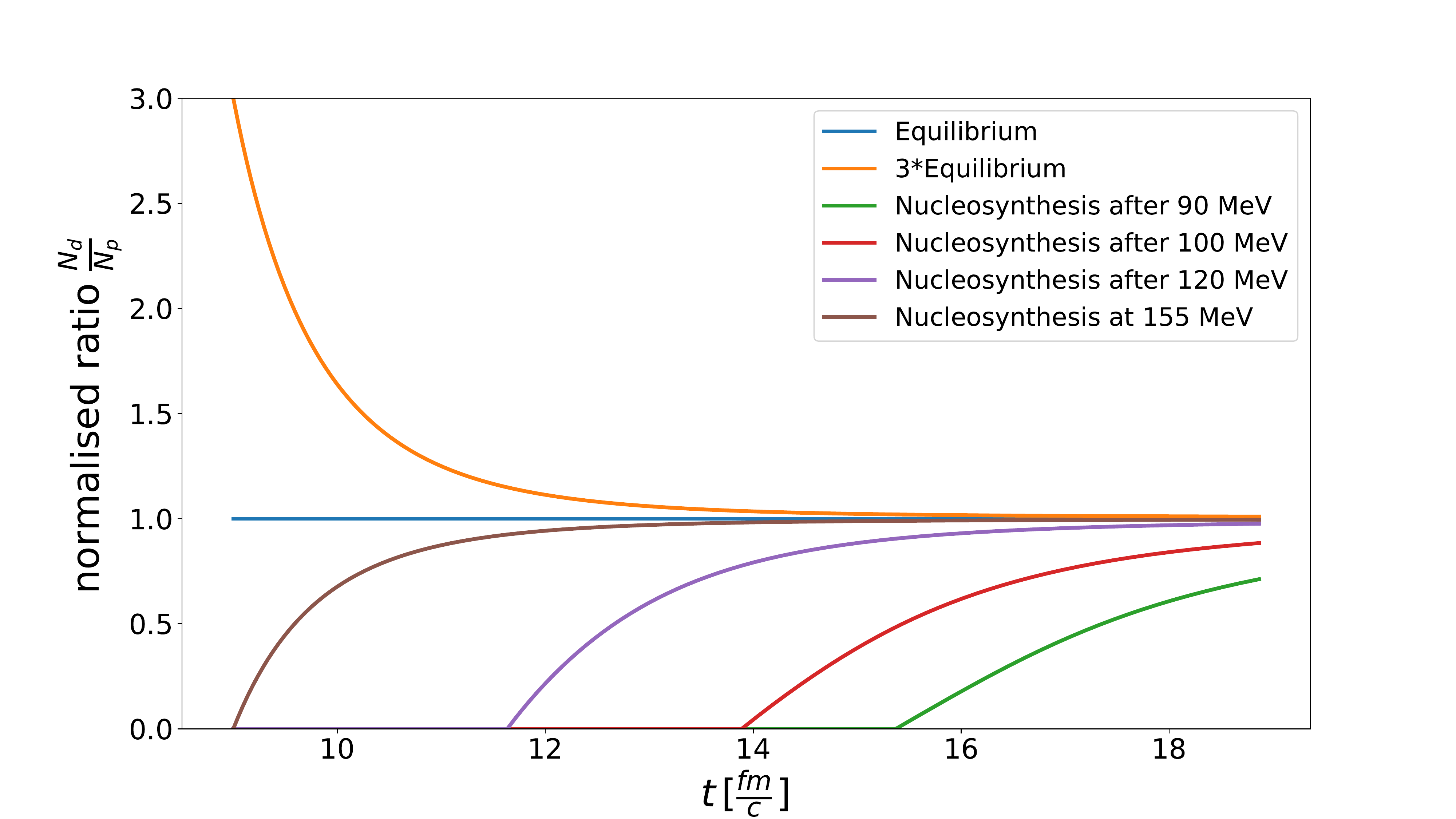}
    \hspace*{\fill}%

    \caption{
      Ratio of deuterons to protons normalised to the same ratio when starting from equilibrium at $T_{\rm ch}=155 \MeV$ for different initial conditions.
    }
    \label{fig:ratiosdp}
  \end{center}
\end{figure}

Next we investigate the equilibration time of light nuclei starting from an equilibrated hadron gas without clusters. We also demonstrate that cluster (re-)equilibration is robust with respect to the thermal value when the recombination reactions start not immediately after the chemical freeze-out but at lower temperatures (nucleosynthesis starts only at $T=90, 100, 120$ and $155 \MeV$). For this purpose we look at the behavior of the deuteron to proton ratio obtained in these different scenarios as shown in  \cref{fig:ratiosdp} as function of time. Qualitatively the same results were found for the other light nuclei, which are indeed approaching the equilibrium value slightly faster.

It is notable
that one reaches the ($T=155\MeV$)-equilibrium ratio even if one starts with the potential regeneration reactions only at 120\MeV. For lower temperatures like 100\MeV or 90\MeV only 88\proz or 70\proz of the equilibrium value is reached. The typical kinetic freeze-out is expected to take place between $100 \MeV < T_{\rm kin} < 120 \MeV$~\cite{ALICE:2015wav,Motornenko:2019jha} in central collisions, so that one is tempted to take the conclusion that the freeze-out conditions for the light nuclei are close to the ($T=155\MeV$)-equilibrium and dictated by the principle of PCE.

\begin{figure}[htb]
  \begin{center}
    \hspace*{\fill}%
    \includegraphics[width=0.95\columnwidth,clip=true]{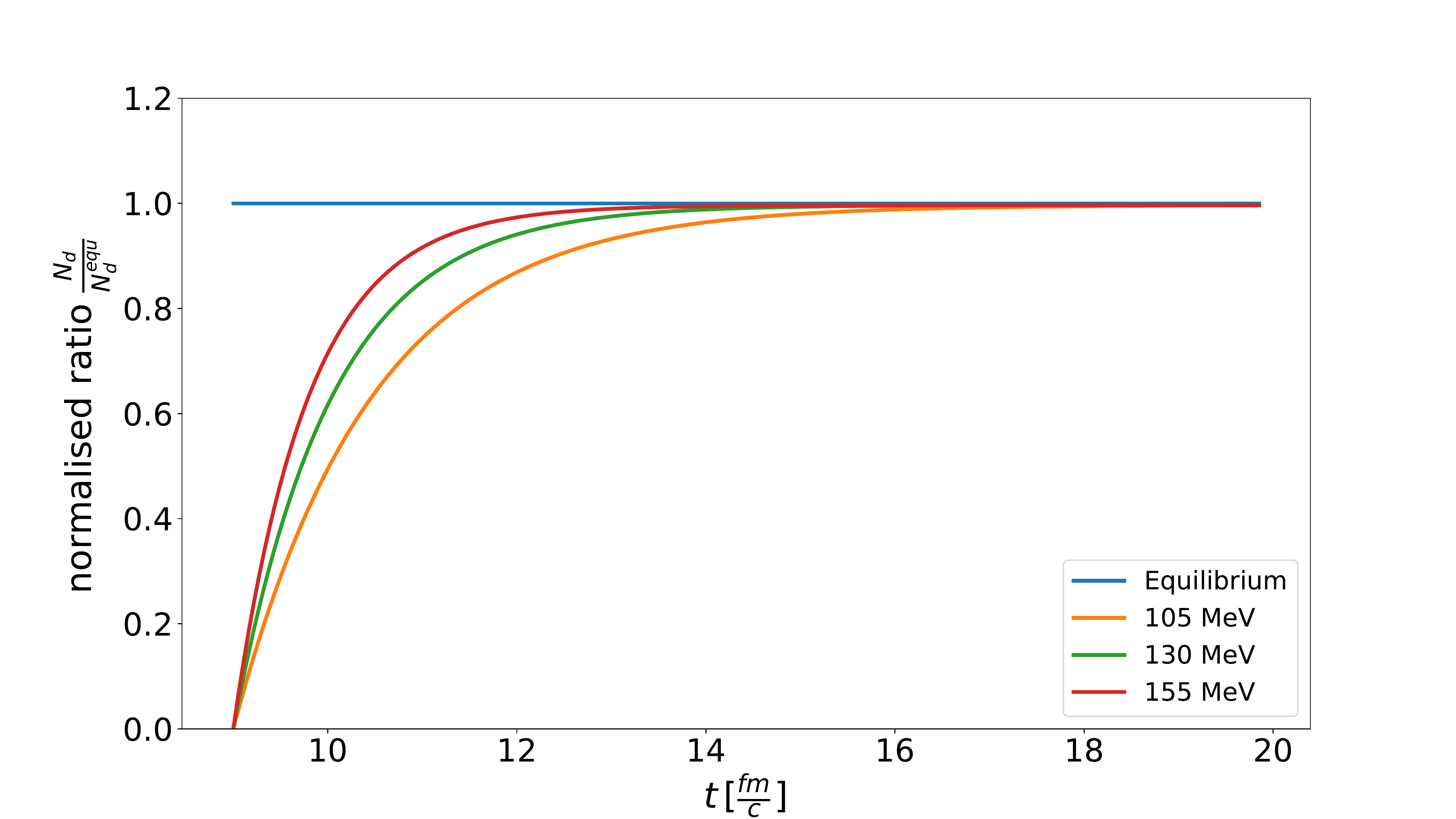}
    \hspace*{\fill}%

    \caption{
      Time evolution of the number of deuterons normalized to the equilibrium value with $V=4000 \fm^{3}$ and starting from zero initial yield for three different temperatures: $T=T_{\rm ch}=155 \MeV$ (orange), $T=130 \MeV$ (green) and $T=105 \MeV$ (red)
    }
    \label{fig:equratio}
  \end{center}
\end{figure}

\begin{figure}[htb]
  \begin{center}
    \hspace*{\fill}%
    \includegraphics[width=0.95\columnwidth,clip=true]{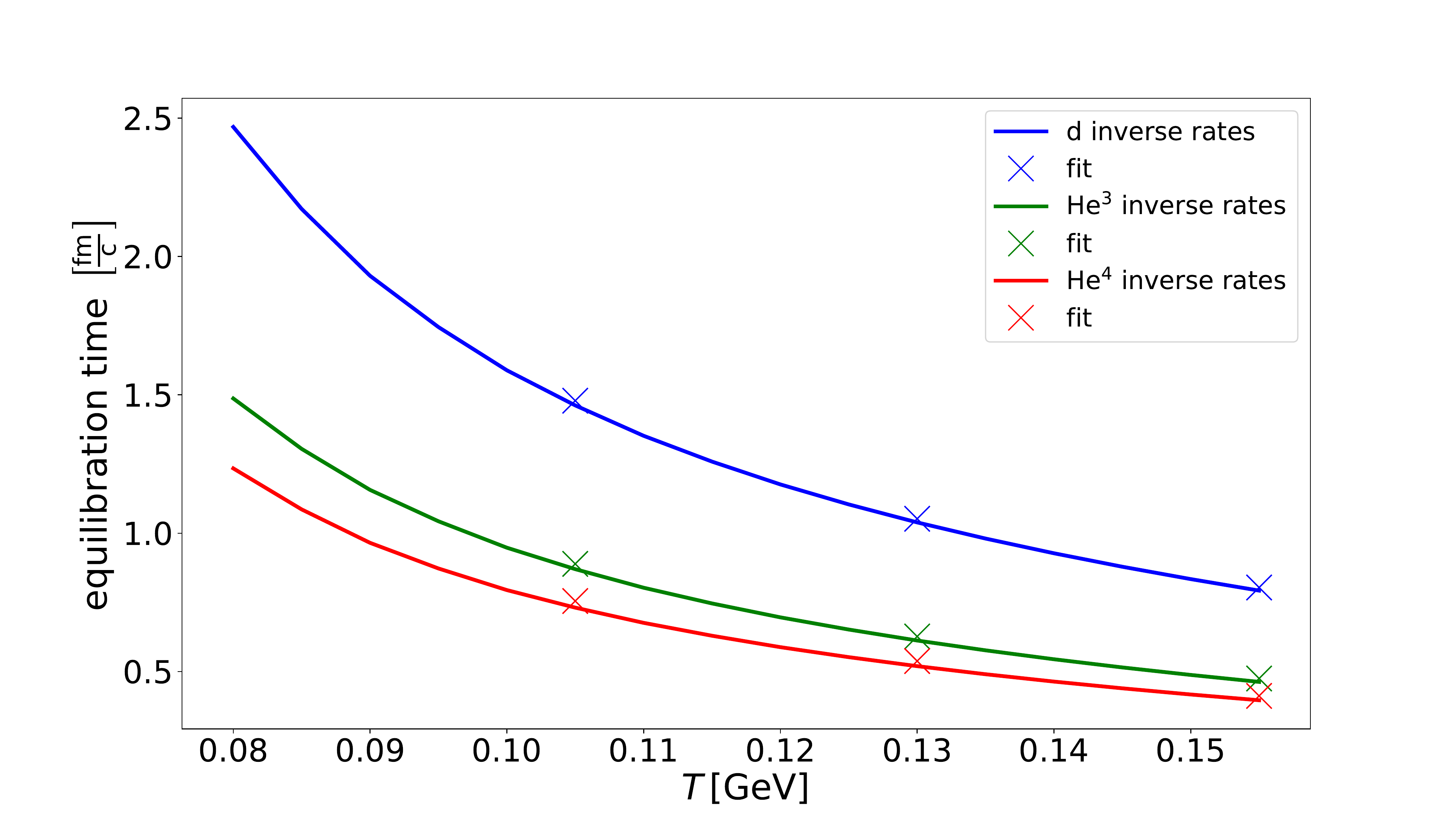}
    \hspace*{\fill}%

    \caption{
      Equilibration time of different light nuclei for different temperatures in a box ($V=4000 \fm^{3}$). The solid line represents the inverse of the sum of the rates and the dots are obtained by exponential fits to curves from \cref{fig:equratio} and alike.
    }
    \label{fig:equ}
  \end{center}
\end{figure}

We also want to stress the connection between the annihilation rates $\alpha$ for each reaction and the equilibration time of a particle species. For fixed volume and a given temperature the hadrons and resonances are in full thermal and chemical equilibrium and the abundancies of the light nuclei are set to zero. As a particular example we show the normalised ratio of deuterons as a function of time. The results are shown in \cref{fig:equratio}. 
By performing an exponential fit to these different temperature branches, one can extract the equilibration time of the deuterons. For \cref{fig:equ} this has also be done for ${\rm He}^3$ and ${\rm He}^4$. By comparing these results to the inverse of the sum of the rates, those are in perfect agreement with the formula \cref{eq:4} for a light nucleus A,
\begin{align}
\frac{1}{\tau^{eq}_{A}}= \sum_{x=\pi,\ki,\ka} \al{\nA+x}{aN+x} N_{x} = \alpha_{A}.
\label{eq:4}
\end{align}

It follows from \cref{fig:equ} that the equilibration time of the light nuclei are very short and in the range 0.5-1.5\fmc. 


So far with our network, the amount of stable hadrons is strictly conserved. Finally, we now would like to study the potential role of baryon-antibaryon annihilations on the abundances of light nuclei.
The possible relevance of baryon annihilation has been raised since a long time \cite{Rapp:2000gy, Greiner:2000tu}, and has recently gained considerable attention
at SPS, RHIC, and LHC energies \cite{Steinheimer:2012rd, Becattini:2012xb,Seifert:2017oyb, Seifert:2018bwl,Savchuk:2021aog, Garcia-Montero:2021haa}.
These processes can be easily implemented in our network. Let us for instance take the process~$N + \overline{N} \rightleftharpoons  5 \pi$ \cite{Xu:2018jff}. We take $5 \pi$ as a mean value of annihilating pions, it could in principle vary in a range 3-7$\pi$ \cite{Noronha-Hostler:2007fzh}.
This reaction is not part of the original PCE but can be incorporated within the rate equations approach as follows:
\begin{align}
  \begin{split}
\frac{\dd N_N}{\dd t} &\,\pluseq \al{N + \overline{N}}{5 \pi} ( - N_{N} N_{\overline{N}} + \cR{N\overline{N}}{\pi^5} \, N_{\pi}^{5}) \\
\frac{\dd N_{\overline{N}}}{\dd t} &\,\pluseq  \al{N + \overline{N}}{5 \pi} ( - N_{N} N_{\overline{N}} + \cR{N\overline{N}}{\pi^5} \, N_{\pi}^{5}) \\
\frac{\dd N_{\pi}}{\dd t} &\,\pluseq 5 \al{N + \overline{N}}{5 \pi} ( N_{N} N_{\overline{N}} - \cR{N\overline{N}}{\pi^5} \, N_{\pi}^{5}) \ . \label{eq:52}
\end{split}
\end{align}

\begin{figure}[htb]
  \begin{center}
    \hspace*{\fill}%
    \includegraphics[width=0.95\columnwidth,clip=true]{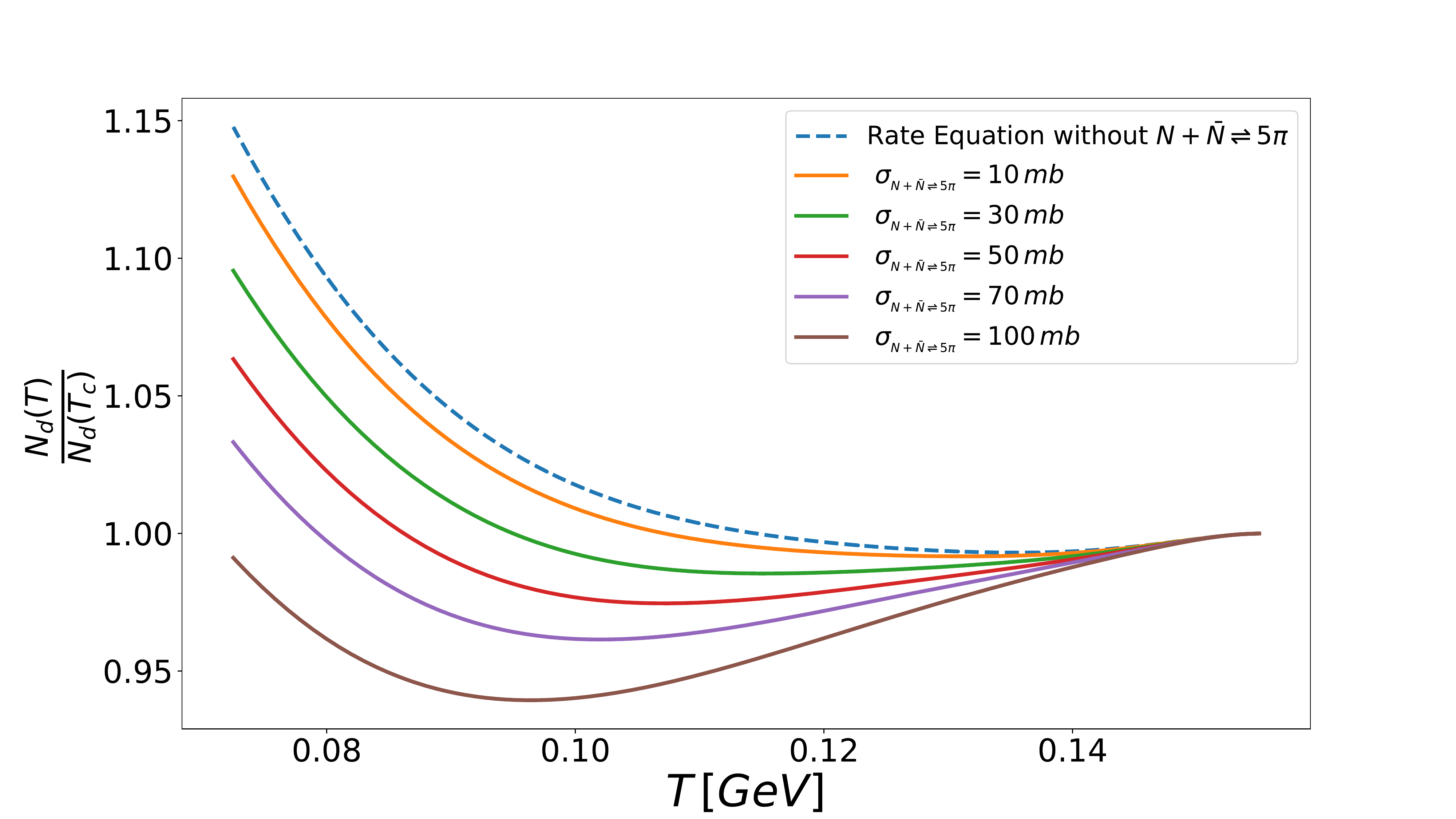}
    \hspace*{\fill}%

    \caption{
    Deuteron yield at various temperatures normalised to the equilibrium value at $T_{\rm ch}=155 \MeV$ in a system described by the network of rate equations with (strong) baryon-antibaryon annihilation reactions $N + \overline{N}$ for various values of the corresponding cross section.
    }
    \label{fig:5Pi}
  \end{center}
\end{figure}

In principle, the cross section for these reactions can be taken from the data on inelastic $p + \overline{p}$ scattering. 
Here, however, we vary its thermal averaged value in a broad range spanning 0-100\mb. Typically 50\mb \cite{Xu:2018jff} is the appropriate value for the inelastic/annihilation cross section.
The corresponding effect of the deuteron yields at decreasing temperatures $T<T_{\rm ch}$ is depicted in \cref{fig:5Pi}.
The results indicate that baryon annihilation leads to a suppression of the deuteron number at intermediate stages of the expansion, but these modifications are small ($5\proz$) at the kinetic freeze-out at $T=100 \MeV$. The influence of the baryon annihilation is somewhat stronger for the heavier nuclei. Here, the modifications are up $25\%$, but still in the range of the experimental uncertainty.
Hence, our main conclusions obtained without incorporating $N\overline{N}$ annihilation are unaffected. 
%


In summary, we have shown that the light nuclei abundances equilibrate quickly towards their partial chemical equilibrium yields in an expanding hadronic environment. Because of the PCE, the abundance of the light nuclei show only a minor temperature dependence when decreasing the decoupling temperature. This explains why the late stage coalescence results in similar light nuclei yields as a thermal analysis with $T_{\rm{ch}}=155\MeV$, albeit coalescence happens at a much lower temperature. 

To this aim, we employed a set of 23 coupled rate equations among stable hadrons, resonances, and light (strange)(anti-)nuclei. The corresponding chemical equilibration times are are found to be very short, i.e.~between 0.5\fmc close to the chemical freeze-out and 1-1.5\fmc at lower temperatures around $T_{\rm kin} \sim 100\MeV$    
of the (potential sequential) kinetic freeze-out of the hadronic fireball. Consequently, the abundances stay very close to their ''initial'' values given by the statistical model at the chemical freeze-out and may be accurately described by a Saha equation~\cite{Vovchenko:2019aoz}.

This implies that the light nuclei (e.g.~deuterons) do not necessarily have to be formed at the chemical freeze-out:
Even if the nucleosynthesis only starts at temperatures as low as $120 \MeV$, the statistical model values are essentially reached in the final state. If the nucleosynthesis would start only at $90\MeV$ one still obtains 70\proz of the initial equilibrium yield.

We have also considered the effect of annihilations between nucleons and anti-nucleons and found that it only leads to a small modification in the total yields of the light nuclei.
As such, our study emphasizes the validity of the (partial) chemical equilibrium scenario for the expanding hadronic fireball created in heavy-ion collisions.

One may be even come to the conclusion that the remarkable agreement of the experimental yields of light nuclei with the yields obtained within the statistical hadronization description \cite{Andronic:2017pug} at the temperature of $T_{\rm ch} \approx 155\MeV$ leads strong support to the principle of the partial chemical equilibrium, and hence the validity of the Saha equation.

All results given are based on the law of mass action. If there are (strong) annihilation reactions, (strong) generation processes must also take place according to principle of detailed balance. It does, however, not tell how the (re-)generation processes are realized microscopically, but they have to occur in time. Of course, the understanding of the microscopic process of the generating processes is a formidable quantum mechanical task, which is outside the scope of the present study.


\section*{Acknowledgements}
V.V. thanks V. Koch and D. Oliinychenko for fruitful discussions.
K.G. was supported by the Bundesministerium f\"ur Bildung und
Forschung (BMBF), grant No.~3313040033.
We jointly acknowledge support by the Deutsche Forschungsgemeinschaft (DFG) through the CRC-TR 211 'Strong-interaction matter under extreme conditions'.
V.V. acknowledges the support through the
Feodor Lynen Program of the Alexander von Humboldt
foundation, the U.S. Department of Energy, 
Office of Science, Office of Nuclear Physics, under contract number 
DE-AC02-05CH11231231, and within the framework of the
Beam Energy Scan Theory (BEST) Topical Collaboration.

\appendix



\bibliography{references}%

\end{document}